\documentclass[a4paper,10pt]{sigwebnewsletter}
\usepackage{subfigure}
\usepackage{booktabs}
\usepackage{enumitem}

\title{Turing Number: How Far Are You to A. M. Turing Award?}

\author{Feng Xia$^{1,2}$, Jiaying Liu$^{1}$, Jing Ren$^{1}$, Wei Wang$^{3}$, Xiangjie Kong$^{4}$ \\
$^{1}$ Dalian University of Technology, China\\
$^{2}$ Federation University Australia, Australia \\
$^{3}$ University of Macau, China \\
$^{4}$ Zhejiang University	of Technology, China}

\newsletterQuarter{Autumn}
\newsletterYear{2020}

\begin{abstract}
The ACM A.M. Turing Award is commonly acknowledged as the highest distinction in the realm of computer science. Since 1960s, it has been awarded to computer scientists who made outstanding contributions. The significance of this award is far-reaching to the laureates as well as their research teams. However, unlike the Nobel Prize that has been extensively investigated, little research has been done to explore this most important award. To this end, we propose the Turing Number ($TN$) index to measure how far a specific scholar is to this award. Inspired by previous works on Erdos Number and Bacon Number, this index is defined as the shortest path between a given scholar to any Turing Award Laureate. Experimental results suggest that $TN$ can reflect the closeness of collaboration between scholars and Turing Award Laureates. With the correlation analysis between $TN$ and metrics from the bibliometric-level and network-level, we demonstrate that $TN$ has the potential of reflecting a scholar's academic influence and reputation.
\end{abstract}

\begin{document}

\maketitle

\section{Itroduction}
As the "Nobel Prize of Computing", the ACM A.M. Turing Award is named for the great British mathematician, Alan M. Turing. More than fifty years have passed since the ACM awarded its first Turing Award to Alan Perlis. As of 2019, over 70 scholars from around the world have received the Turing Award~\cite{fisher2017turing,kong2019evolution}. However, the research of Turing Award seems like a virgin territory compared with the Nobel Prize. For example, considerable effort has been done in terms of understanding the characteristics of Nobel Prize Laureates as well as predicting future winners~\cite{ren2019api,kong2020gene,bai2020quantifying}. However, when you try to search any study on the Turing Award from Google Scholar, you may get disappointed because little related research or papers has been done or published. We believe that analyzing the highest award of computing is beneficial to highlight the significant contribution of the Turing Award Laureates, look ahead to the future of computing, and help motivate other computer scientists to dream and create~\cite{hanson2017celebrating,zhang2020data}.

One of the promising research directions on exploring Turing Award is the behavior dynamics of its laureates. Similar research has been done in terms of Nobel Award Laureates~\cite{schlagberger2016institutions}, which examines the possibility of predicting the Nobel Award Laureates. At the same time, network-based approaches have been extensively utilized to explore the collaboration behaviors of scientists~\cite{liu2019shifu2,liu2020web}. Inspired by previous work on Erdos Number and Bacon Number, this study tries to answer an interesting question in terms of Turing Award: How far a computer scientist is to the great scientist Turing? In other words, how to measure the shortest path between a given scholar and Turing. In the era of Turing, scientists are not as collaborative as nowadays and Turing had no collaborator when publishing papers. Therefore, an alternative question is proposed: How far a computer scientist is to the Turing Award Laureates? Specifically, we propose "Turing Number" index ($TN$) which measures the shortest path between a given scholar to any Turing Award Laureate.

As $TN$ measures the closeness of collaboration between scholars and Turing Award Laureates, we assume that it can be used to measure the influence of scholars in the Turing collaboration network. To verify this hypothesis, we design experiments from the bibliometric-level and network-level, respectively. From the perspective of bibliometric, we calculate the productivity (e.g., number of papers) and impact (e.g., citations and h-index) for both scholars and countries, and examine the relationships between these metrics and $TN$. From the perspective of network structure, we use different indices such as degree centrality, closeness centrality, and betweenness centrality to measure the importance of different $TN$ scholars in Turing collaboration networks. Experimental results suggest that $TN$ can not only reflect the closeness of collaboration between scholars and Turing Award Laureates, but also measure the influence of scholars. Our contributions of this article are summarized as follows:
\begin{itemize}
	\item We propose a new way to construct the scientific collboration network centering around the Turing Award Laureates.
	\item We define a new metric $TN$ to quantify the distance between scholars and Turing Award Laureates.
	\item We provide detailed correlation analysis of various metrics from the bibliometric-level and network-level, with the aim of exploring the relationship between $TN$ and scholar's academic influence and reputation.
\end{itemize}

\section{Social Network Analysis}

\textbf{Dataset} To construct the scientific collaboration network, the DBLP\footnote{https://dblp.org/} dataset is used in this work and we select the computer scientists by 2020. All related information consist of scholar's name, paper title, and their publication years. The geographical location information and citation information of each scholar is acquired by matching the author information with Aminer dataset~\cite{tang2008arnetminer} to get the institution information and citation relationships. Then, we use the Google Map API\footnote{https://cloud.google.com/maps-platform/} to get the latitude and longitude of the institutions, as well as the region information. Authors' name disambiguation process is conducted based on previous corresponding work~\cite{sinatra2016quantifying}. The Turing Award Laureates are identified by name manually. For institutions' name disambiguation, we use the regular expression to search and match the institution information. Meanwhile, institutions with missing records are excluded. For scholars with more than one institution, we use the institution information by the investigating year. The final collaboration network includes 53,531 scholars including 65 Turing Award Laureates and their collaborators from 119 countries and regions. In this collaboration network, nodes represent scholars and edges represent collaboration relationships between scholars. Based on the constructed network, we calculate $TN$ for each scholar as follows.

\textbf{Turing Number (\emph{TN})}
Turing number refers to the network distance of an author to the Turing Award Laureates. A Turing number is defined as the minimum of the shortest path of each author to all Turing Award Laureates. To be assigned a Turing number, someone must be a coauthor of a research paper with another person who has a finite Turing number. A Turing Award laureate has a Turing number of zero. Anybody else's Turing number is $k + 1$ where $k$ is the lowest Turing number of any coauthor. Based on its definition, a smaller $TN$ denotes that the scholar is closer to the Turing Award.

Following previous conceptualizations, we explore $TN$ from two aspects: bibliometric-level and network-level. The bibliometric indicators we used include the number of publications, citations, and h-index. Table~\ref{tab:network} lists a detailed description of network-level indices used in this work.

Furthermore, for better understanding the special position of the Turing Award Laureates in the network and eliminating geographical bias, we construct a null model for comparison. The null model which is constructed by randomly selecting equal number of scholars, i.e., 65 as the Turing Award Laureates.

\begin{table}%
	\caption{Description of network-level indices.}
	\label{tab:network}
	\begin{minipage}{\columnwidth}
		\begin{center}

			\begin{tabular}{lp{5cm}p{4cm}}
				\hline
				\textbf{Index} & \textbf{Description} & \textbf{Calculation formula} \\
				\hline
				Degree Centrality    &  the number of links incident upon a node   & $C_{D}(v)=deg(v)$ \\
				Closeness Centrality & the reciprocal of the sum of the length of the shortest paths between the node and all other nodes in the graph &$C_{C}(v)= \frac{|V|-1}{\sum_{i\neq v}d_{vi}}$\\
				Betweenness centrality 	&the number of times a node occurs on the shortest path between other nodes & $B_{C}(v)=\sum_{s\neq v\neq t\in V}\frac{\sigma_{st}(v)}{\sigma_{st}}	$\\
				Eigenvector Centrality&an algorithm for measuring the transmission impact and connectivity between nodes&the calculation process can be found in~\cite{bonacich1987power}\\
				Load Centrality & the fraction of all shortest paths that pass through that node&the calculation process can be found in~\cite{newman2001scientific,goh2001universal}\\
				\hline
			\end{tabular}
		\end{center}
		\bigskip\centering
	\end{minipage}
\end{table}%

\section{Results and Discussions}
\subsection{Bibliometric-level analysis}
\begin{figure}
	\centering
	\subfigure[Scholar distribution of $TN$]{
		\label{figure:1-a}
		\includegraphics[width=0.485\textwidth]{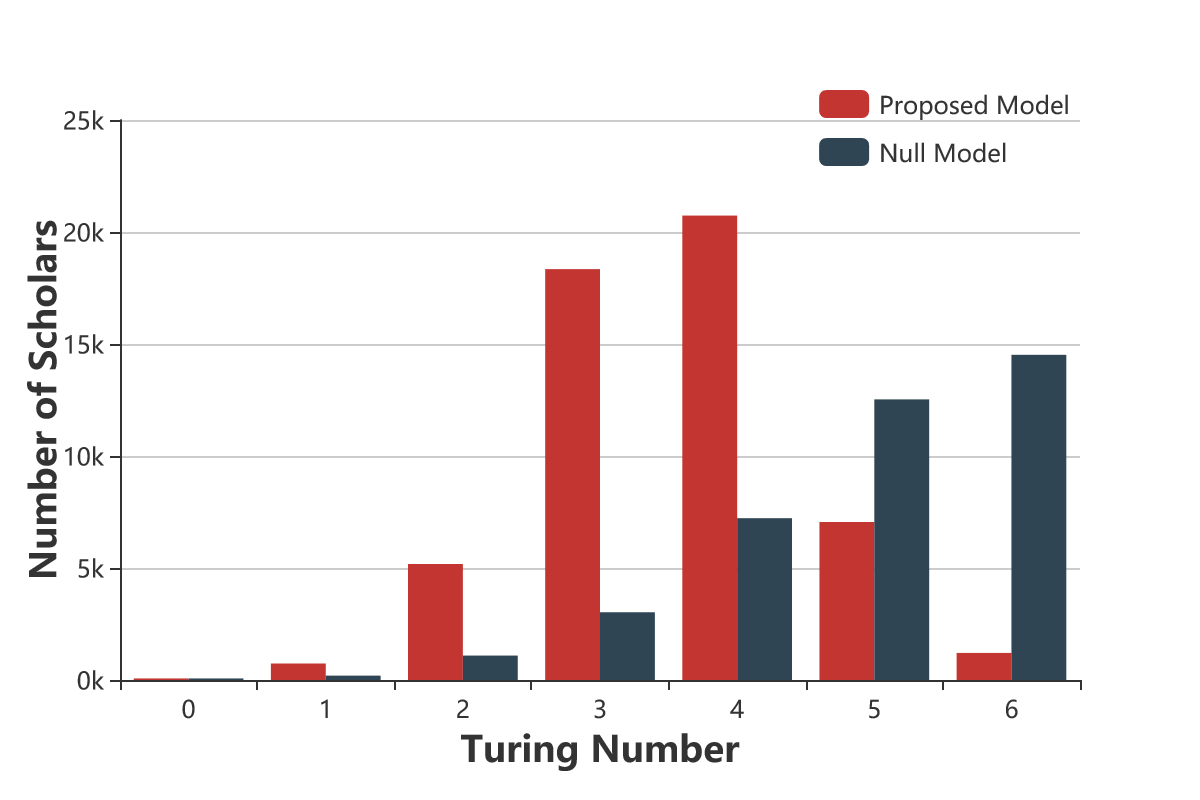}}
	\subfigure[Geographic distribution of $TN$ ]{
		\label{figure:1-b}
		\includegraphics[width=0.485\textwidth]{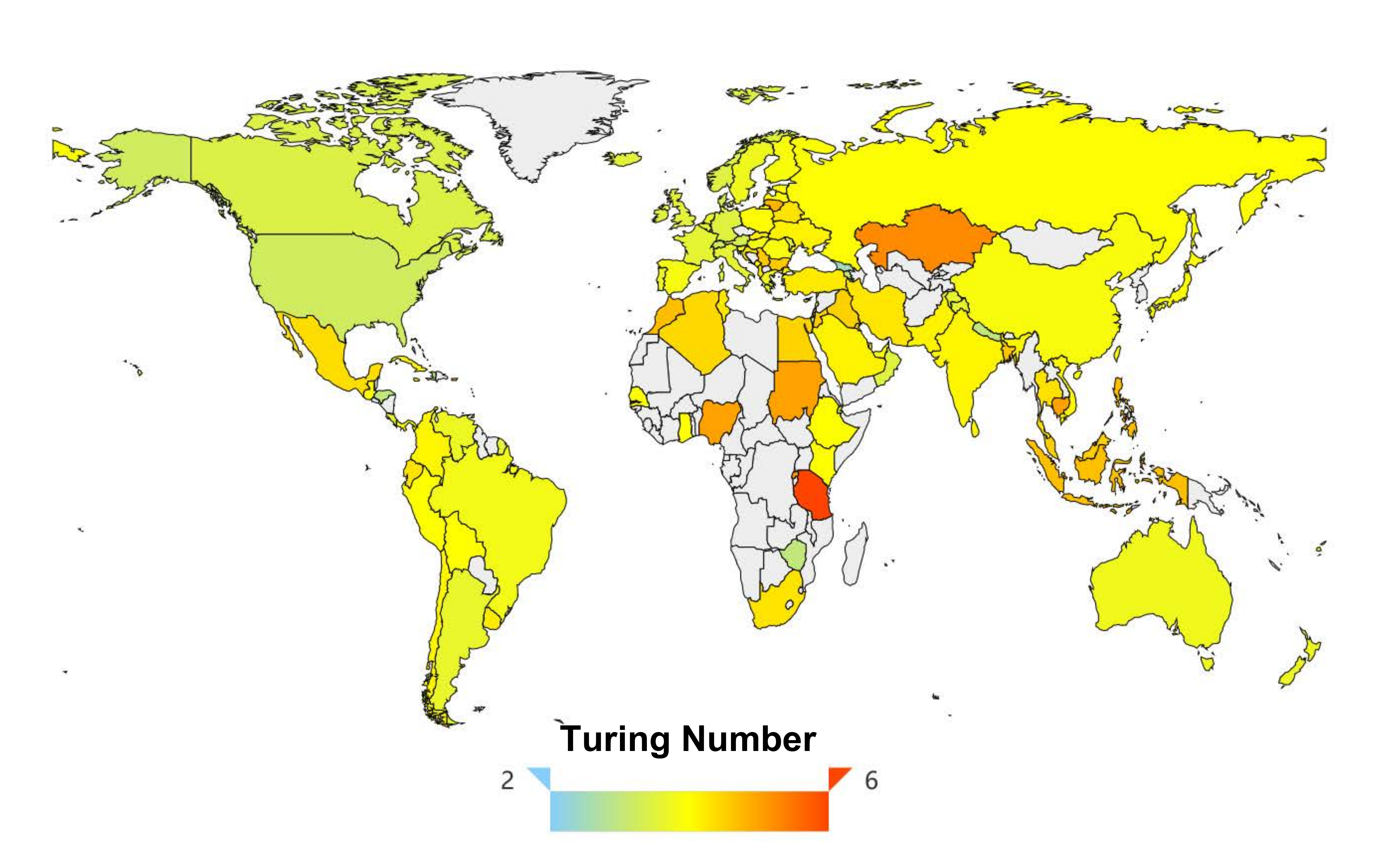}}
	\caption{Distribution of $TN$.}
	\label{figure:1}
\end{figure}

To study how far a computer scientist is to the Turing Award, we first explore the distribution of $TN$ over all investigated scholars and countries (see Fig.~\ref{figure:1}). Fig.~\ref{figure:1-a} shows most scholars have a $TN$ of 3. Meanwhile, more than 90\% scholars' $TN$ is between 2 and 5. It indicates that it takes 3 to 5 jumps for a scholar to get connected with a Turing Award Laureate. This observation confirms previous finding on six-degree separation~\cite{Watts1998CollectiveDO}, which means scholars can reach each other by a relatively small number of connections. However, since the $TN$ is measured by the distance from a scholar to the Turing Award laureate community, it is much smaller. Note that there are few people whose $TN$ is more than 5, which means that the Turing Award laureate collaboration network is well connected. Meanwhile, as shown in Fig.~\ref{figure:1-a}, the $TN$ number in the null model is much higher than that of real Turing Award Laureates network. Fig.~\ref{figure:1-b} shows the geographical distribution of $TN$. We can see that the U.S. achieves a lower $TN$. The Asia and Africa scholars have higher value of $TN$ than scholars from Europe and America.

\begin{figure}
	\centering
	\includegraphics[width=0.7\textwidth]{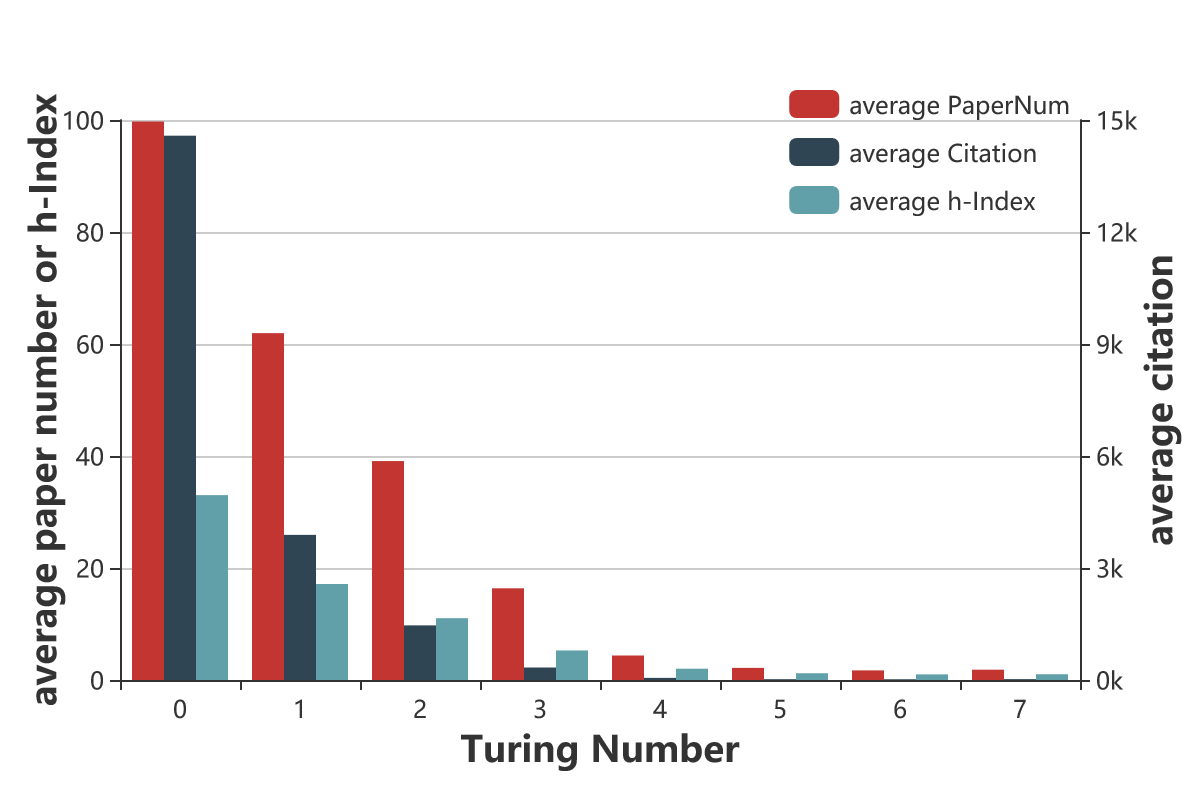}\\
	\caption{Scholars with different $TN$ in terms of the number of publications, citations, and h-index.}
	\label{fig:2}
\end{figure}

Figure~\ref{fig:2} shows the statistics of scholars with different $TN$ in terms of number of publications, citations, and h-index. It shows that with the increase of $TN$, the number of publications declines accordingly. The average number of publications of scholars with $TN=1$ is 61.64 while that of scholars with $TN=5$ is 2.17. Scholars' average impact in terms of citations and h-index also declines significantly with the increase of $TN$. For instance, the average h-index of scholars with $TN=1$ is 17.14 while that of scholars with $TN=5$ is only 1.21. All experimental results verify our hypothesis that $TN$ can reflect the academic influence of scholars.

Table~\ref{tab:one} shows the statistics of $TN$ in the top 11 countries in terms of the number of scholars. It is acknowledged that The United States is advanced in the field of computer science. Not surprisingly, it has the best performance in terms of number of scholars, citations, and $TN$. Influential countries in computer science area such as Germany, France and Canada are also high on the list. All of them have high impact in terms of average h-index and low $TN$. Besides that, emerging countries such as China and India are also on the list. The study by Nature Index has shown that China's international papers have increased to 24\% in 2016, which merely lags behind Germany, the United States and the United Kingdom~\cite{Phillips2017ACL}. So we can believe that $TN$ is a good measure to reflect the national scientific research level.

\begin{table}%
	\caption{Statistics of $TN$ in top countries.}
	\label{tab:one}
	\begin{minipage}{\columnwidth}
		\begin{center}
			\begin{tabular}{llllll}
				\toprule
				\textbf{Country} & \textbf{Scholars} & \textbf{Papers} & \textbf{Citations} & \textbf{h-index}  & \textbf{TN}\\
				\toprule
				United States & 16,791 & 15.40 & 657.56 & 5.58&3.24\\
				China	&3,438	&9.84	&161.63&2.92&3.89\\
				Germany &2,477&14.48&	299.55	&4.63&3.48\\
				France	&2,279	&13.58&	248.30	&4.42&3.58\\
				Canada	&1,960&	14.93&	537.91&4.78&3.44\\
				United Kingdom	&1,906	&14.67	&416.52&5.08 &3.53\\
				Japan	&1,378&	10.68&	106.41&	2.75&3.98\\
				Brazil&1,374	&	9.14	&	84.86	&2.58&3.81\\
				South Korea&1,272	&7.83	&109.57&	2.80&4.07\\
				Italy	&1,095	&24.57&	436.10	&6.45&3.58\\
				India	&1,078	&15.62&		227.45	&4.52&3.86\\
				
				\bottomrule
			\end{tabular}
		\end{center}
		\bigskip\centering
	\end{minipage}
\end{table}%

To further understand the association between $TN$ and conventional bibliometric indicators, we perform correlation analysis between the two sets of variables. The results are shown in Table~\ref{tab:3}. We can observe from this table that the correlation coefficients between $TN$ and bibliometric indicators are negative. Besides that, the correlation between each index and $TN$ is always significant. H-index has the highest correlation with $TN$ in comparison with citations and number of publications. Overall, $TN$ is related to bibliometric indicators. Scholars with lower $TN$ have better academic performance, which demonstrates the feasibility of utilizing $TN$ to measure scholars' impact.

\begin{table}%
	\caption{Results of correlation analysis.}
	\label{tab:3}
	\begin{minipage}{\columnwidth}
		\begin{center}
			\begin{tabular}{cccc}
				\toprule
				\textbf{Correlation analysis method} & \textbf{No. of Publications} & \textbf{No. of Citations} & \textbf{H-index} \\
				\toprule
				Pearson   &-0.330***&    -0.258***& -0.440***\\
				Spearman	&  -0.463 ***	& -0.450***	& -0.490***\\
				Kendall & -0.381***&    -0.393***&  -0.408***\\
				\bottomrule
				{$*p<0.1; **p < 0.05; ***p< 0.01$}
				
			\end{tabular}
		\end{center}
		\bigskip\centering
	\end{minipage}
\end{table}%

\subsection{Network-level analysis}
\begin{figure}
	\centering
	\includegraphics[width=0.8\textwidth]{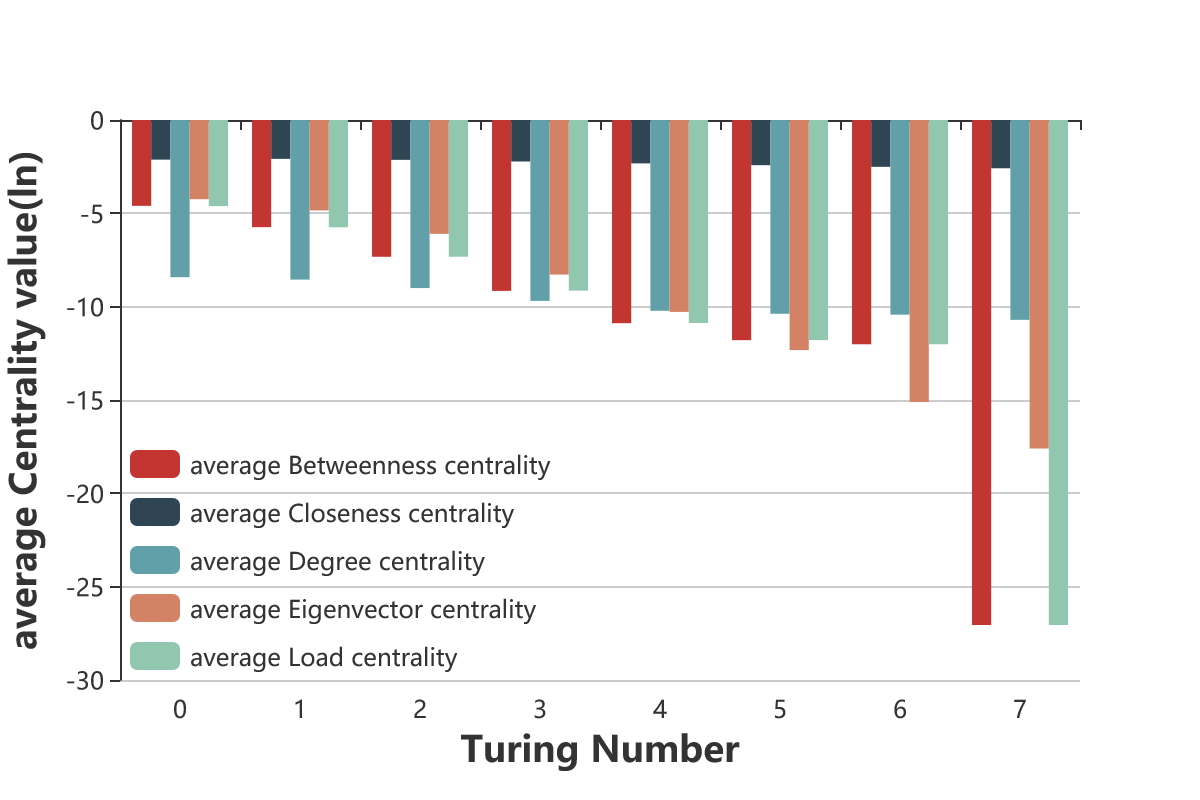}\\
	\caption{Distribution of $TN$ conditional to network centrality measures. }
	\label{fig:3}
\end{figure}

The analysis of network level indicators is helpful to understand the evolution and development of Turing collaboration network and find the key nodes in the network. From the network structure perspective, we explore the distribution of $TN$ conditional to network centrality measures mentioned in Table~\ref{tab:network}. The result is shown in Fig.~\ref{fig:3}. Specifically, we take the $\ln$ of each index. We can see that with the increase of $TN$, the value of each indicator gradually becomes smaller. For example, for betweenness centrality, it ranges from $0.01$ to $1.78e^{-12}$ while $TN$ changes from 0 to 7. Overall, the result shows that scholars with lower $TN$ are more important in the network in terms of degree centrality, between centrality, closeness centrality, eigenvector centrality, and load centrality. $TN$ is not only related to bibliometric indicators but also network centrality indicators. Scholars with lower $TN$ have more important positions in the collaboration network, which demonstrates the feasibility of utilizing $TN$ to measure scholars' importance from the network structure level.

\section{Conclusion}
Turing Award is the highest honor in the field of computer science. In this article, we investigated the potential rules of Turing Award collaboration network and explored the relationship and distance between an ordinary scholar and Turing Award Laureates. Our results showed that most scholars are close to the Turing Award Laureates within three steps in the network. The correlation analysis between this indicator and traditional bibliometric indicators suggested that $TN$ is meaningful for scholars' impact evaluation. Besides, our comparison of $TN$ among different countries showed the feasibility of utilizing this indicator to evaluate the influence of a given country. Furthermore, $TN$ can be regarded as a measure of the collaboration with top computer scientists around the world. It can somehow reflect the reputation of a given scholar. By examining the relationships between $TN$ and network centrality indicators, we concluded that scholars with lower value of $TN$ generally have more important positions in the collaboration network.

\section*{ACKNOWLEDGMENTS}

This work is partially supported by National Natural Science Foundation of China under Grant No. 61872054 and the Fundamental Research Funds for the Central Universities (DUT19LAB23).

\bibliographystyle{sigwebnewsletter}
\nocite{*}
\bibliography{TN}

\begin{biography}
\textit{Feng Xia} received the BSc and Ph.D. degrees from Zhejiang University, Hangzhou, China. He is currently an Associate Professor and Discipline Leader in School of Engineering, IT and Physical Sciences, Federation University Australia, and on leave from School of Software, Dalian University of Technology, China, where he is a Full Professor. Dr. Xia has published 2 books and over 300 scientific papers in international journals and conferences. His research interests include data science, social computing, and systems engineering. He is a Senior Member of IEEE and ACM. (\textit{Corresponding author; email: f.xia@ieee.org})

\textit{Jiaying Liu} received the BSc degree in Software Engineering from Dalian University of Technology, China, in 2016. She is currently working toward the PhD degree in School of Software, Dalian University of Technology, China. Her research interests include data science, big scholarly data, and social network analysis.

\textit{Jing Ren} received the Bachelor’s degree from Huaqiao University, China, in 2018, and the Master degree from Dalian University of Technology, China, in 2020. She is currently a Research Fellow in the School of Software, Dalian University of Technology, China. Her research interests include data science, social computing, and graph learning.

\textit{Wei Wang} is a Research Fellow at University of Macau, Macau SAR. He received PhD degree in Software Engineering from Dalian University of Technology in 2018. His research interests include computational social science, data mining, internet of things, and artificial intelligence. He has authored/co-authored over 50 scientific papers in international journals and conferences. He is a member of IEEE, ACM, and CCF.

\textit{Xiangjie Kong} received the B.Sc. and Ph.D. degrees from Zhejiang University, Hangzhou, China. He is currently a Professor with  College of Computer Science and Technology, Zhejiang University of Technology, China. He has published over 100 scientific papers in international journals and conferences. His research interests include network science, data science, and computational social science. He is a Senior Member of the IEEE and CCF and is a member of ACM.
\end{biography}

\end{document}